\input epsf.tex    

\documentstyle[12pt]{article}

\newcommand{\bmat}{\left(\begin{array}}
\newcommand{\emat}{\end{array}\right)}

\def\yzero{\smash{\hbox{$y\kern-4pt\raise1pt\hbox{${}^\circ$}$}}}

\def\beq{\begin{equation}}
\def\eeq{\end{equation}}
\def\beqa{\begin{eqnarray}}
\def\eeqa{\end{eqnarray}}

\def\-{\hphantom{-}}
\def\ov{\overline}
\def\s2{\frac{1}{2}}

\def\beq{\begin{equation}}
\def\eeq{\end{equation}}
\def\beqa{\begin{eqnarray}}
\def\eeqa{\end{eqnarray}}

\def\IF{\relax{\rm I\kern-.18em F}}
\def\II{\relax{\rm I\kern-.18em I}}

\def\cp{{\cal P}}
\def\IC{\bf C}
\def\IZ{\bf Z}
\def\IR{\bf R}

\def\IS{\bf S}
\def\IP{\bf P}
\def\IT{\bf T}
\def\IM{\bf M}
\def\IX{\bf X}
\def\z2z2{$\IC^3/(\IZ_2\times\IZ_2)$}

\def\NN{{\cal N}}
\def\Dsl{\,\raise.15ex\hbox{/}\mkern-13.5mu D} 

 \def\cp#1{\relax\ifmmode {\IP\kern-2pt{}_{#1}}\else $\IP\kern-2pt{}_{#1}$\=fi}

\begin{document}

\makeatletter \@addtoreset{equation}{section} \makeatother
\renewcommand{\theequation}{\thesection.\arabic{equation}}

\pagestyle{empty}
\vspace*{.5in}
\rightline{FTUAM-03/13, IFT-UAM/CSIC-03-25}
\rightline{\tt hep-th/0307156}
\vspace{1.5cm}
 
\begin{center}
\LARGE{\bf M5-brane geometries, T-duality and fluxes \\[10mm]}

\medskip

\large{Juan F. G. Cascales$^1$, Angel M. Uranga$^2$} \\
{\normalsize {\em $^1$ Departamento de F\'{\i}sica Te\'orica C-XI \\
and Instituto de F\'{\i}sica Te\'orica, C-XVI \\
Universidad Aut\'onoma de Madrid \\
Cantoblanco, 28049 Madrid, Spain \\
{\tt juan.garcia@uam.es}\\
$^2$ IMAFF and \\
 Instituto de F\'{\i}sica Te\'orica, C-XVI \\
Universidad Aut\'onoma de Madrid \\
Cantoblanco, 28049 Madrid, Spain \\
{\tt angel.uranga@uam.es} \\[2mm]}}

\end{center}

\smallskip

\begin{center}
\begin{minipage}[h]{14.5cm}
{\small
We describe a duality relation between configurations of M5-branes in M-theory and 
type IIB theory on Taub-NUT geometries with NSNS and RR 3-form field strength fluxes. 
The flux parameters are controlled by the angles between the M5-brane and the 
(T)duality directions. For one M5-brane, the duality leads to a family of 
supersymmetric flux configurations which interpolates between imaginary self-dual 
fluxes and fluxes similar to the Polchinski-Strassler kind. For multiple M5-branes, 
the IIB configurations are related to fluxes for twisted sector fields in orbifolds. 
The dual M5-brane picture also provides a geometric interpretation for several 
properties of flux configurations (like the supersymmetry conditions, their 
contribution to tadpoles, etc), and for many non-trivial effects in the IIB side. Among 
the latter, the dielectric effect for probe D3-branes is dual to the recombination of 
probe M5-branes with background ones; also, a picture of a decay channel for 
non-supersymmetric fluxes is suggested.}

\end{minipage}
\end{center}

\newpage                                                        


\setcounter{page}{1} \pagestyle{plain}
\renewcommand{\thefootnote}{\arabic{footnote}}
\setcounter{footnote}{0}

\section{Introduction}

String and M-theory backgrounds with field strength fluxes for various $p$-form gauge 
fields have been shown to lead to interesting features 
\cite{fluxes,beckers,gkp,ferrara,chiral}. For instance, scalar potentials with 
non-trivial minima leading to moduli stabilization, supergravity descriptions in terms 
of warped geometries, and (possibly partial) breaking of supersymmetry. These 
configurations are presently under intense study from diverse points of view. In this 
paper we would like to present a new approach to their analysis in terms of a new dual 
picture. 

We describe a duality between configurations of M-theory M5-branes and type IIB string 
theory on Taub-NUT geometries with a background of NSNS and RR 3-form field strength 
fluxes. Data of the flux configuration, and many of its properties, are encoded in 
simple properties of the geometry of the dual M5-brane configuration. 

The duality is a generalization of the relation between parallel M5-branes in M-theory 
and Taub-NUT geometries in IIB theory, which we review in Section \ref{dualityone}. In 
Section \ref{dualitytwo} we present our duality and show that M5-branes at angles with 
respect to the duality directions are related to Taub-NUT geometries with a non-trivial 
background of 3-form fluxes, controlled by the dual angle parameters (sections 
\ref{intuitive} and \ref{detailed}). 

These configurations are interesting for several reasons. Different properties of the 
IIB backgrounds are encoded in the M5-brane geometry, for instance flux quantization 
conditions (section \ref{quantization}), and charges induced by the flux background 
(section \ref{induced}). Moreover, in Section \ref{supersymmetry} we argue that the IIB 
duals to single stacks of M5-branes preserve 16 supersymmetries, and provide a family 
of backgrounds that interpolate between imaginary self-dual fluxes with constant 
dilaton and more general supersymmetric flux configurations with varying dilaton, 
reminiscent of those in \cite{ps}. 

In section \ref{multiple} we discuss multiple M5-brane stacks, for which the M-theory 
dual provides a description of twisted fluxes on $\IC^2/\IZ_N$ orbifold singularities. 
Again some features like moduli stabilization or the supersymmetry of the configuration 
are explained using the M5-brane geometry (section \ref{moduli}). The M-theory picture 
can be also used to suggest decay mechanisms for non-supersymmetric fluxes (section 
\ref{recombination}), and the stabilization by fluxes of unstable non-Calabi-Yau 
geometries (section \ref{tachyons}). 

In section \ref{probes} we discuss the M-theory picture of diverse effects when one 
introduces D3-brane probes in the IIB flux background, for instance, the appearance of 
4d $\NN=2$ soft masses (section \ref{masses}), and Myers dielectric effect \cite{myers} 
(section \ref{dielectric}). A prominent role is played by the process of recombination 
of intersecting M5-branes. Finally, in section \ref{foursusy} we discuss 
generalizations of the above duality to threefold geometries related to the conifold. 
Section \ref{conclusions} contains our final comments.

\section{M5-brane vs. Taub-NUT}
\label{dualityone}

In this section we review the geometry of Taub-NUT and its duality relation with 
M-theory M5-branes.

\subsection{A few facts on Taub-NUT}
        
The $N$-center Taub-NUT metric is (see e.g. \cite{sen} for a useful reference)
\beqa
ds^2 & = & H(\vec{x})^{-1}(dy+\omega_i dx^i)^2 \, + \, H(\vec{x}) 
d\vec{x}^{\,2} \nonumber \\
H(\vec{x}) & = & 1+\sum_{a=1}^N H_a(\vec{x}) \quad ; \quad 
H_a(\vec{x})= \frac{1}{|\vec{x}-\vec{x}_a|} \nonumber \\
d\omega & = & *_{3d}\, dH(\vec{x}) 
\label{taubnut}
\eeqa
where $i$ runs over three indices and $\vec{x}=(x^i)$. Also $\omega=\omega_i dx^i$
and $*_{3d}$ denotes Hodge duality in the $\IR^3$ parametrized by $\vec{x}$.

The geometry corresponds to an $\IS^1$ fibration over a base $\IR^3$ (parametrized by 
$\vec{x}$). The fiber has constant asymptotic radius (fixed to unity in the above 
expression) and degenerates to zero radius over the locations $\vec{x}_a$, known as 
centers of the geometry. The $\IS^1$ bundle over a 2-sphere $(\IS^2)_k$ in 
$\IR^3$ surrounding $k$ of these centers has Chern class $k$, namely
\beqa
\int_{(\IS^2)_k} d\omega=k
\label{tntwist}
\eeqa
so is a Hopf bundle over a two-sphere surrounding each center. 

The geometry contains $N-1$ homologically independent non-trivial compact 
2-cycles. A non-trivial 2-cycle $\Sigma_{ab}$ can be obtained by fibering 
the $\IS^1$ fiber over the segment in $\IR^3$ joining the centers 
$\vec{x}_a$ and $\vec{x}_b$. Out of these $\Sigma_{ab}$, only $N-1$ are linearly 
independent, a simple basis is provided by the cycles $\Sigma_{a,a+1}$, 
for $a=1,\ldots, N-1$.

The geometry also contains $N$ cohomologically independent normalizable 2-forms, 
\beqa
\Omega_a=d\chi_a \quad ; \quad \chi_a=H^{-1}H_a(dy+\omega)-\omega_a
\eeqa
where $\omega_a$ is defined by $d\omega_a=*_{3d}d H_a(\vec{x})$.
With suitable normalization they obey the orthonormality condition $\int_{TN}
\Omega_a \wedge \Omega_b= \delta_{ab}$.

The form $\Omega_a$ has support localized near the center $\vec{x}_a$. It 
is useful to introduce linear combinations $\Omega=\sum_{a=1}^N \Omega_a$ 
and $\Omega_{ab}=(\Omega_a-\Omega_b)/2$. The latter are Poincare dual to the 
2-cycles $\Sigma_{ab}$ (with suitable orientation). In the limit of infinite 
asymptotic radius, the geometry corresponds to an $N$-center ALE geometry, 
namely a (generically blown-up) $\IC^2/\IZ_N$ orbifold. In this limit, the form 
$\Omega$ becomes non-normalizable, and can be regarded as a constant 2-form 
inherited from the covering $\IC^2$. The forms 
\beqa
\Omega_k\, =\,\frac 1N\,  \sum_{a=1}^N \, e^{2\pi i\, ka/N} \, \Omega_{a}
\eeqa
(with $a$ understood modulo $N$) belong to the $k^{th}$ twisted sector of 
the orbifold. We will hence denote the forms $\Omega\equiv\Omega_{0}$, 
$\Omega_{k\neq 0}$ as untwisted and twisted, respectively.

For a one-center metric, $H=1+1/r$ with $r=|\vec{x}|$, and the harmonic 2-form is
\beqa
\Omega = - d(H^{-1})(dy+\omega)-H^{-1} d\omega
\label{twoforms}
\eeqa

\subsection{The duality relation}
\label{duality}

We begin by reviewing the duality relation between M-theory configurations of 
parallel M5-branes (with two transverse coordinates compactified on a two-torus)
with type IIB on a Taub-NUT metric background. This relation is usually phrased
as a T-duality between IIA NS5-branes and IIB on Taub-NUT geometries 
\cite{oovafa}, but the former formulation makes the geometry of forthcoming
configurations more transparent.

Consider a set of $N$ M5-branes with world-volume along the directions 012345, and 
sitting at a point in the remaining ones, 678910. Let 6,10 be compactified on a 
two-torus that we momentarily take square for simplicity. 

The relation to IIB on Taub-NUT is most easily established using the 
supergravity description of the M-theory configuration. It is convenient to 
use the solution for M5-branes smeared in the directions 6, 10 (see 
\cite{tong} for a discussion of the localized M5-brane solution and T-duality). 
It reads 
\beqa
ds^2_{11d} & = & H_5(\vec{x})^{-1/3} \, \, ds_{012345}^2 \,
\, + \, H_5(\vec{x})^{2/3}\, \, ds_{678910}^2 
\label{mfive}
\eeqa
where $H_5(\vec{x})= 1+\sum_a \frac{1}{|\vec{x}-\vec{x}_a|}$, 
and $\vec{x}_a$ denote the M5-brane locations in 789. For simplicity we 
consider the case of coincident centers $\vec{x}_a=0$, for which
$H_{5}=1+\frac{N}{r}$, with $r=|\vec{x}|$.

Moreover, the M5-branes are magnetically charged under the M-theory 3-form.
Hence there is a 3-form background $C_3=\omega\, dx^6\, dx^{10}$ (wedge 
products are implicit throughout the paper), with $d\omega=*_{3d}dH_5$.
Upon reduction to IIA in the direction 10, using the standard ansatz
\beqa
ds^2 & = & e^{4\phi/3}\, (\,dx^{10}\,+\,A_M\,dx^M\,)^2 \, +\, e^{-2\phi/3}\,
g^{IIA}_{MN} \,dx^M \,dx^N
\eeqa
we obtain the IIA background
\beqa
ds_{IIA}^2 & = & ds_{012345}^2 + H_5\, ds_{6789}^2 \nonumber \\
e^{2\phi} &  = & H_5 \nonumber \\
B^{NS} & = & \omega\, dx^6
\eeqa
Application of T-duality formulae in appendix \ref{trules}, leads to the
type IIB background fields
\beqa
\tau_{IIB} & = & a+ie^{-\phi}= i \nonumber\\
ds_{IIB}^2 & = & ds_{012345}^2 + H_5^{-1}(r) (dx^6 +\omega_i dx^i)^2 + H_5(r) 
d\vec{x}^{\, 2}
\eeqa

This corresponds to a purely metric background, corresponding to a Taub-NUT
geometry (\ref{taubnut}) with $N$ coincident centers. The IIB complex coupling
takes the value $\tau=i$ because we took equal radii for the 6, 10 circles. Applying 
the dualities to the background with general radii leads to $\tau=iR_6/R_{10}$.

The general mapping of parameters between both configurations is as follows. 
The location of the $a^{th}$ M5-brane in 7, 8, 9 maps to the position $\vec{x}_a$ of 
the $a^{th}$ Taub-NUT center in the base $\IR^3$. The location of the $a^{th}$ M5-brane 
in 6, 10 is mapped to the component of the NSNS and RR 2-form fields along 
$\Omega_a$. Namely, the positions (normalized w.r.t. the total radius, 
i.e. $\phi^i=x^i/R_i$) correspond to the coefficients $\phi_a^6$, $\phi_a^{10}$ 
in the expansion
\beqa
B_{NSNS} = \sum_a \phi_a^6\, \Omega_a \quad ; \quad
B_{RR} = \sum_a \phi_a^{10}\, \Omega_a 
\label{posit}
\eeqa
See \cite{kls} for further discussion. The bottomline of this section is 
that simultaneous shrinking of the directions 6, 10 in the M5-brane
configuration leads to IIB theory on a Taub-NUT geometry.

\section{M5-brane geometries and fluxes}
\label{dualitytwo}

In this section we consider a generalization of the above duality. It is a 
duality between configurations of M5-branes at angles
and Taub-NUT geometries with 3-form field strength fluxes. Intuitively, the
amount of rotation of a given M5-brane with respect to the 4, 5 coordinates
will map to the amount of NSNS and RR flux turned on along the corresponding
harmonic 2-form in the dual Taub-NUT geometry. Thus, many features of fluxes
are easily geometrized into configurations of M5-branes.

\subsection{Single M5-brane}
\label{single}

\subsubsection{Intuitive T-duality}
\label{intuitive}

\begin{figure}[]
\begin{center}
\centering
\epsfysize=3.5cm
\leavevmode
\epsfbox{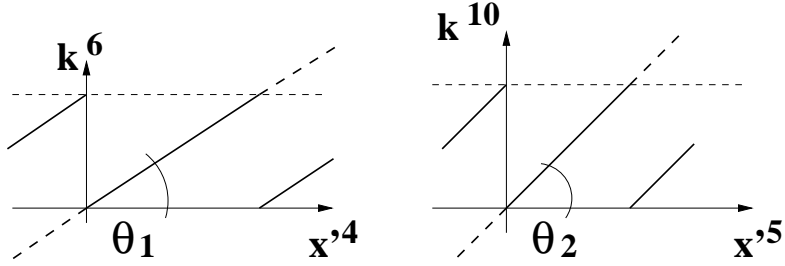}
\end{center}
\caption[]{\small Configuration of M5-brane at angles (with respect to 
the duality directions).}
\label{angles}
\end{figure}

Let us start by considering an intuitive explanation of the basic duality
we explore in the present paper. Let us regard the M-theory spacetime as 
the product of a $\IM_4$ (spanned by 0123), times $\IR\times \IS^1$ 
(spanned by 46), times $\IR\times \IS^1$ (spanned by 510), times an
$\IR^3$ (spanned by 789). Consider a single M5 brane with
volume spanning $\IM_4$ times a real line in each of the $\IR\times \IS^1$
factors. We denote by $\theta_1$, $\theta_2$ the angle between these lines
and the directions 4 and 5, respectively. See figure \ref{angles}.

Let us shrink the directions 6, 10 and obtain the corresponding dual type IIB
configuration. In order to do that, it is useful to regard the original M-theory
configuration as that of one M5-brane spanning 0, 1, 2, 3, 4, 5 and with a
non-trivial profile for the world-volume scalars $\phi^6$, $\phi^{10}$ 
that encode the location of the M5-brane in the directions 6, 10, of the 
form
\beqa
\phi^6\,=\, \tan \theta_1\, x^4 \quad ; \quad \phi^{10}\,=\,\tan \theta_2\, x^5
\label{lines}
\eeqa
Under
the duality, we obtain the IIB theory on a Taub-NUT geometry. Recall that 
the scalars are mapped to the components of $B_{NSNS}$, $B_{RR}$ along 
$\Omega$. The linear variation of these scalars with $x^4$, $x^5$ then 
implies that there are non-trivial NSNS and RR 3-form field strength fluxes
turned on the Taub-NUT, roughly of the form
\beqa
H_3\,=\, \tan \theta_1\, \Omega\, dx^4 \quad ;\quad
F_3\,=\, \tan \theta_2\, \Omega\, dx^5
\label{simpleflux}
\eeqa
(Normalization can easily be fixed from flux quantization conditions, see section 
\ref{quantization}).

The generalization  to $N$ parallel M5-branes is straightforward, namely 
the IIB dual is given by an $N$-center Taub-NUT space, with fluxes 
(\ref{simpleflux}) along the overall (untwisted) 2-form $\Omega$.

Hence we have found a duality relation between configurations of 3-form field
strength fluxes in Taub-NUT space and the geometry of the M5-brane. The above
analysis is however oversimplified, e.g. it ignores the backreaction of
the fluxes in the geometry. In next section we describe a more careful
application of T-duality, that reproduces these effects.

\subsubsection{Detailed duality relation}
\label{detailed}

\begin{figure}
\begin{center}
\centering
\epsfysize=3.8cm
\leavevmode
\epsfbox{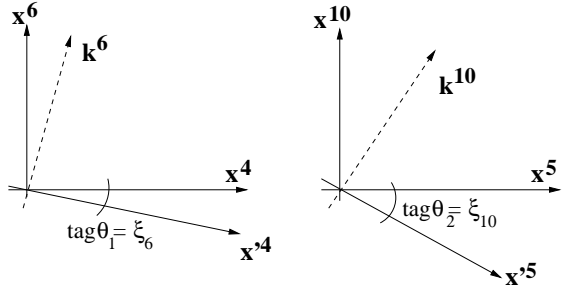}
\end{center}
\caption[]{\small Pictorial depiction of the effect of changing coordinates. 
The directions which are shrunk in the duality are at fixed values of the new 
coordinates $x'^4$, $x'^5$, hence correspond to the Killing directions $k_6$, 
$k_{10}$. Hence we are describing a (T)duality of the M5-brane configuration along 
directions at angles with the latter. The angles $\theta_1$, $\theta_2$ are determined 
by the coordinate change parameters $\xi_6$, $\xi_{10}$. In order to
avoid annoying minus signs in our formulas, we consider clockwise
angles as positive.}
\label{killing}
\end{figure}

Let us derive the above duality relation with an argument closer to that in
section \ref{duality}. Consider the supergravity solution corresponding 
to M5-branes along 012345, in  a background Minkowski metric, 
(\ref{mfive}). In order to introduce the above tilting of the M5-brane, 
let us perform the change of variables
\beqa
x^4\, =\, x'^4\,+\,\xi_6\, x^6 \quad ; \quad
x^5\, =\, x'^5\,+\,\xi_{10}\, x^{10}
\label{tilt}
\eeqa
As shown in figure (\ref{killing}), this implies that the M5-brane is along directions 
at angles with respect to the directions along which we perform the dualities.

The metric becomes 
\beqa
ds^2_{11d} & = & H_5^{-1/3}\, ds_{0123}^2 \, + \, H_5^{2/3}\, ds_{789}^2 
+ H_5^{-1/3}(H_5+\xi_6^2) (dx^6)^2 +  H_5^{-1/3}(H_5+\xi_{10}^2) (dx^{10})^2
\nonumber \\
& + & H_5^{-1/3} (dx'^4)^2 +  H_5^{-1/3} (dx'^5)^2 
+ 2\xi_6\, H_5^{-1/3} dx'^4 dx^6 + 2\xi_{10}\, H_5^{-1/3} dx'^5 dx^{10}
\eeqa

Let us ignore for the moment the effect of the coordinate change in the 
3-form background, and take $C_3=\omega \, dx^6\, dx^{10}$. Its full discussion 
is postponed to section \ref{induced}. Also, in what follows, we drop the primes 
of the new 45 coordinates.

Upon reduction to IIA, we obtain the background
\beqa
e^{4\phi/3} & = & H_5^{-1/3}(H_5+\xi_{10}^2) \nonumber \\
A_5 & = & \frac{\xi_{10}}{H_5+\xi_{10}^2}\nonumber \\
ds_{IIA}^2 & = & H_5^{-1/2} \,(H_5+\xi_{10}^2)^{1/2}\, ds_{0123}^2 \,+
H_5^{1/2}\, (H_5+\xi_{10}^2)^{1/2}\, ds_{789}^2 + \nonumber \\
& + & H_5^{-1/2}\, (H_5+\xi_{10}^2)^{1/2}\, (H_5+\xi_6^2)\, (dx^6)^2 \,+ \,
 2 \xi_6 \,H_5^{-1/2}\, (H_5+\xi_{10}^2)^{1/2}\, dx^4\, dx^6\, + \nonumber \\
& + & H_5^{-1/2}\, (H_5+\xi_{10}^2)^{1/2}\, (dx^4)^2\, +\,
H_5^{1/2}\, (H_5+\xi_{10}^2)^{-1/2}\, (dx^5)^2 
\eeqa
and we have the $2$-form background 
\beqa
B_{NSNS} & = & \omega \, dx^6 
\eeqa
Application of the T-duality rules in appendix \ref{trules} leads to the type IIB 
background \footnote{The metric has a more symmetric form in the Einstein frame, as 
required by S-duality, which amounts to an exchange of the roles of $\xi_6, \xi_{10}$.}

\beqa 
ds_{IIB}^2 & = &  H_5^{-1/2}\, (H_5+\xi_{10}^2)^{1/2}\, ds_{0123}^2\, +\, H_5^{1/2}\, (H_5+\xi_{10}^2)^{1/2}\, (H_5+\xi_6^2)^{-1}\, (dx^4)^2\,+\nonumber \\
& + &  H_5^{1/2}\, (H_5+\xi_{10}^2)^{-1/2}\, (dx^5)^2 +\nonumber \\
& + & H_5^{1/2}\, (H_5+\xi_{10}^2)^{-1/2}\, (H_5+\xi_6^2)^{-1}\, 
(\,dx^6\,+\,\omega_i\, dx^i\,)^2\, +\,  H_5^{1/2}\, (H_5+\xi_{10}^2)^{1/2}\, ds_{789}^2 \,\nonumber \\
e^{-\phi} & = & \left( \frac{H_5+\xi_6^2}{H_5+\xi_{10}^2} \right)^{1/2} \nonumber\\
B_{NSNS} & =- & \frac{\xi_6}{H_5+\xi_6^2} \, (\,dx^6\,+\, \omega_i\, dx^i\,) 
\wedge dx^4
\nonumber \\
B_{RR} & =- & \frac{\xi_{10}}{H_5+\xi_{10}^2} \, (\,dx^6\,+\, \omega_i\, dx^i) 
\wedge dx^5
\eeqa 

Again, more careful discussion of the M-theory 3-form (see section 
\ref{induced}) introduces additional pieces in the above 2-forms.

As discussed above, it corresponds to a (deformed) Taub-NUT background with 
3-form fluxes. Note that the flux densities are controlled by the  
parameters $\xi_6$, $\xi_{10}$. Their field strengths 
\beqa
H_3 & = & -\xi_6 \, [\, d(H_5+\xi_6^2)^{-1} \, (dx^6+\omega)\, +\,
(H_5+\xi_6^2)^{-1}\, d\omega\,] \wedge dx^ 4 \nonumber \\
F_3 & = & -\xi_{10} \, [\, d(H_5+\xi_{10}^2)^{-1} \, (dx^6+\omega)\, +\,
(H_5+\xi_{10}^2)^{-1}\, d\omega] \wedge dx^5
\label{fieldstr}
\eeqa
correspond to harmonic 2-forms in the above deformed geometry, in the following sense. 
As the flux parameters approach zero $\xi_6, \xi_{10}\to 0$ (and the metric
approaches the undeformed Taub-NUT and the dilaton becomes constant), the field
strengths are proportional to the harmonic 2-form (\ref{twoforms}). A 
sketchy description of the configuration is, therefore, a Taub-NUT 
geometry with 3-form fluxes (\ref{simpleflux}). Namely, the tilting of the 
M5-brane with respect to the (T)duality directions in 46,  and 510, turns 
into NSNS and RR 3-form fluxes, respectively, in the dual IIB 
configuration. Further backreaction of the fluxes on the geometry leads to a 
squashing of the latter, which appears at quadratic order in the flux perturbation.

\medskip

The generalization of the duality relation is clear. Type IIB on a Taub-NUT
geometry, with 3-form fluxes \footnote{By this we mean that these would be 
the expressions for the fluxes in the undeformed Taub-NUT metric. The 
backreaction of the flux on the metric subsequently changes the form of 
the flux itself. Hence in this discussion we work in the `flux probe' 
approximation.} 
\beqa
H_3 & = & \Omega \, ( a_1 dx^4 + a_2 dx^5 )\nonumber \\
F_3 & = & \Omega \, ( a_3 dx^4 + a_5 dx^5 )
\label{generalflux}
\eeqa 
is dual to a configuration of M5-branes with volume spanning $\IM_4$ times 
the 2-plane defined by $x^6 =  a_1\, x^4 + a_2\, x^5$, $x^{10}= a_3\, x^4 + a_5\, x^5$.

The generalization to arbitrary complex IIB coupling is similar, by introducing
an arbitrary mixing between the coordinates $6$, $10$ in (\ref{tilt}), via 
a change of variables $x^6=x'^6+\rho_{10}x^{10}$. The full IIB solution may 
be found from the above supergravity background by performing an 
$SL(2,\IR)$ transformation. A sketchy description is that the 
familiar 3-form flux combination $G_3=F_3-\Phi H_3$ (with $\Phi$ the IIB complex 
coupling) plays the role of the 
M5-brane position in the complex plane $z=x^{10}-\Phi x^6$. For simplicity 
we will restrict to the above situation with zero axion in most of the paper.

Hence, we have found a duality relation that provides a geometric interpretation 
for 3-form fluxes in Taub-NUT geometries. In the following we discuss the
geometric interpretation of diverse properties of fluxes. For simplicity, we
center on fluxes of the form (\ref{simpleflux}), rather than 
(\ref{generalflux}).

\subsubsection{Flux quantization}
\label{quantization}

Consider a 1 center Taub-NUT geometry and take the directions 4, 5
to be compactified on a rectangular two-torus, so that 
$x^4$, $x^5$ have lengths $R_4$, $R_5$. Let us introduce the 1-forms 
$d\phi^4=dx^4/R_4$, $d\phi^5=dx^5/R_5$, with period 1 over the non-trivial cycles 
in this two-torus. In this compact setup, consistency of the configuration requires the 
3-form fluxes $H_3$, $F_3$ to be quantized. At linear order (i.e. in the 'flux
probe' approximation) this can be phrased as 
\beqa
H_3\, =\, k_6 \,\Omega \, d\phi^4\, +\, k_6'\, \Omega\, d\phi^5
\eeqa
with $k_6,k_6'\in\IZ$, and analogously for $F_3$ (with coefficients $k_{10}'$, $k_{10}$).
Namely the fluxes must define {\em integer} cohomology classes.

The quantization of fluxes has a very natural interpretation in the dual 
M-theory configuration. Consider for simplicity fluxes $H_3=k_6 \Omega d\phi^4$, 
$F_3=k_{10} \Omega d\phi^5$. Recalling the interpretation of the flux coefficients 
as dual positions (\ref{posit}), we have
\beqa
\frac{\partial \phi^6}{\partial x^4}=\frac{k_6}{R_4} \quad ; \quad
\frac{\partial \phi^{10}}{\partial x^5}=\frac{k_{10}}{R_5}
\eeqa
This implies that the total wrapping number of the M5-brane in the direction 
6, as one winds around the direction 4, is an integer 
\beqa
\int_0^{R_4} (\partial \phi^6/\partial x^4) dx^4=k_6
\eeqa
and analogously for the wrapping number of the M5-brane in 10 as one 
moves in 5.

Integrality of the cohomology class of the 3-form fluxes is thus dual to 
the integrality of the homology class of the M5-brane, namely that the 
wrapping numbers along the cycles of the compact directions should be 
integers.

\subsubsection{Induced charges}
\label{induced}

An interesting property of NSNS and RR 3-form fluxes is that their wedge
product acts as a source for the RR 4-form. Namely, the combination of
fluxes carries an induced D3-brane charge. In fact, these are not the
only induced charges carried by our flux configurations. In this section
we describe them and their dual geometric interpretation.

In order to have finite charges we need to consider, as above, compact 4, 5 directions,
which we take of unit length for simplicity.
Consider in the M-theory picture an M5-brane wrapped along one 1-cycle in the
4-6 two-torus and along one 1-cycle in the 5-10 two-torus. The charges of the
configuration are described by the homology class
\beqa
[\Pi] & = & (k_4[a_4]+k_6[a_6])\otimes (k_5[a_5]+k_{10}[a_{10}]) =\\
&=& k_4k_5 [a_4]\otimes [a_5] + k_6 k_{10} [a_6]\otimes [a_{10}] +
k_4 k_{10} [a_4]\otimes [a_{10}] + k_6 k_5 [a_6]\otimes [a_5] \nonumber
\eeqa
where $k_i\in\IZ$ and $[a_i]$ is the 1-cycle along the $i^{th}$ direction.

The charge $k_4 k_5$ is mapped to the Taub-NUT charge (i.e. number of centers) 
in the type IIB dual.
The charge $k_6k_{10}$ should in principle be mapped to a D3-brane 
charge in the IIB dual. Indeed, this is the induced charge due to the 3-form
fluxes, as follows from
\beqa
N_{D3}  & = &
\int_{TN\times \IT^2} H_3\wedge
F_3 = \tan\theta_1\tan\theta_2\ \int_{TN}\Omega\wedge\Omega
 \int_{\IT^2} \, dx^4\,  dx^5\, = \nonumber \\ & = &\, k_4 k_5 \tan\theta_1 \tan\theta_2\, = \,
k_6\, k_{10} 
\label{naived3}
\eeqa
Hence we have found a geometric interpretation for the 4-form charge carried by 
3-form fluxes.

\medskip

There remains to interpret the charges $k_4 k_{10}$ and $k_6k_5$. They correspond to 
charges of D5-branes spanning 46 and NS5-branes spanning 56, respectively. Their 
presence can be understood from the central charge formula for the configuration. 
General results (see \cite{op} or section 3.1. in \cite{pinned}) imply that an 
$N$-center Taub-NUT geometry with a $p$-form background asymptotically along $x^6$ 
develops an induced charge of the corresponding magnetic object. In our context, a 
Taub-NUT charge $N=k_4k_5$ in the presence of a NSNS 2-form asymptoting to 
$\xi_6 dx^4 dx^6 $ develops $Q$ units of induced charge of NS5-brane along 012356, 
with 
\beqa
Q=\xi_6 k_4k_5=\tan\theta_1 k_4k_5=k_6k_5,
\eeqa
as required. Analogously for the D5-brane charge induced by the RR 2-form background.

In the following we rederive these results. Indeed, all the induced charges are 
correctly reproduced once we take into account the change of the M-theory 3-form 
background under the coordinate change (\ref{tilt}), ignored in section 
(\ref{detailed}) for simplicity.

After the coordinate reparametrization (\ref{tilt}), the 11d background 3-form becomes
\beqa
C_3 & = & \omega \, dx^6\, dx^{10} - \xi_6 \omega\, dx^4\, dx^{10} - 
\xi_{10} \omega\, dx^6\, dx^5 + \xi_6 \xi_{10}\omega \, dx^4 \, dx^5
\eeqa
Upon reduction to IIA, we obtain the $p$-form background
\beqa
B_{NSNS} & = & \omega \, dx^6 - \xi_6\,  \omega \, dx^4 \nonumber \\
C_3 & = & -\xi_{10}\, \omega \, dx^6\, dx^5 + \xi_6\, \xi_{10}\, \omega
\, dx^4\, dx^5
\eeqa
Further T-duality along the direction 6 leads to 
\beqa
B_{NSNS} & = &- \frac{\xi_6}{H_5+\xi_6^2} \, (\,dx^6\,+\, \omega_i\, dx^i\,) 
\wedge dx^4 \, - \, \xi_6 \,\omega \wedge dx^4 \nonumber \\
B_{RR} & = & -\frac{\xi_{10}}{H_5+\xi_{10}^2} \, (\,dx^6\,+\, \omega_i\, dx^i) 
\wedge dx^5 \, + \, \xi_{10}\, \omega \wedge dx^5 \nonumber \\
C_4 & = & \xi_{10} \left (\xi_6+\frac{\xi_6}{H_5+\xi_{6}^{2}} 
\right ) \omega \wedge dx^4 \wedge dx^5 \wedge dx^6 
\eeqa
As we now show, the additional pieces in the $p$-form background contain
the additional induced charges discussed above. We sketch their computation, 
using expressions correct to the corresponding order in the fluxes (linear for the 
NS5-, D5-brane charges, quadratic for the D3-brane charge). The NS5- and 
D5-brane 
charges can be obtained by integration of the NSNS resp. RR field 
strength 3-forms over the 3-cycles of the form $\IS^2$ times the direction
$x^4$ resp. $x^5$, where the $\IS^2$ is a 2-sphere in the $\IR^3$ base of 
Taub-NUT, surrounding the centers. This integral vanishes for the first 
piece of the field strengths (\ref{fieldstr}), since they are harmonic and 
the 3-cycle is homologically trivial. Using (\ref{tntwist}), the integral 
of the new piece however gives 
\beqa
\int_{\IS^2\times (\IS^1)_4} \xi_6\, d\omega\, dx^4\, =\, 
k_4\, k_5\, \xi_6\, =\, k_5\, k_6
\eeqa
for the NSNS field, and analogously for the RR field. Hence we recover the 
correct charges.

The induced D3-brane charge is computed similarly. The charge arises from 
integrating the 5-form flux over the 45 two-torus times a large $\IS^3$ in 
Taub-NUT. The latter is obtained from the $\IS^1$ fibration over an $\IS^2$
in the base. Recalling the modification of the 5-form field strength due to 
Chern-Simons couplings,
\beqa
{\tilde F}_5 = dC_4 -\frac{1}{2} B_{RR}\wedge H_3 + \frac{1}{2} F_3\wedge B_{NSNS}
\eeqa
we have
\beqa
\int_{\IS^3\times \IT^2} {\tilde F}_5 = \int_{TN \times \IT^2} d{\tilde F}_5 =
\int_{TN\times \IT^2} H_3\wedge F_3 
\label{intflux}
\eeqa
namely the familiar contribution to the tadpole for the 4-form. In our background, the 
3-form fluxes have the structure 
\beqa
H_3=-\xi_6\Omega\wedge dx^4-\xi_6d\omega \wedge dx^4 \quad ; \quad  
F_3=-\xi_{10}\Omega\wedge dx^5+\xi_{10} d\omega \wedge dx^5
\eeqa
Out of the four contributions to $H_3\wedge F_3$, the piece proportional to $d\omega 
\wedge d\omega$ does not contribute to the integral in (\ref{intflux}), and the two 
pieces proportional to $\Omega \wedge d\omega$ cancel each other. The induced D3-brane 
charge arises from the piece proportional to $\Omega\wedge \Omega$ leading to a total 
charge of $k_4k_5\xi_6\xi_{10}=k_6k_{10}$, as in the naive discussion (\ref{naived3}). 
Hence our flux configuration correctly accounts for all induced charges of 
the system. Alternatively, the M5-brane homology class provides a simple geometric 
interpretation for them.

\medskip

In the above configuration the homology charges, $q_{ij}$, coefficients of the terms 
$[a_i]\otimes [a_j]$ in $[\Pi]$, satisfy the quadratic constraint 
$q_{45}q_{610}=q_{410}q_{65}$. This follows from the factorized form of the wrapped 
2-cycle, and implies that the system preserves 1/2 supersymmetry. It is natural to 
wonder about M5-brane configurations were the homology charges $q_{ij}$ do {\rm not} 
satisfy a quadratic constraint. As discussed in \cite{rabadan}, they correspond to 
M5-branes wrapped on non-factorizable holomorphic 2-cycles, which preserve 1/4 of the 
supersymmetries. These configurations are easily obtained from recombination of 
factorizable M5-branes intersecting at angles not defining an $SU(2)$ rotation. The 
IIB dual of such systems is briefly discussed in section \ref{recombination}.

\subsubsection{Supersymmetry}
\label{supersymmetry}

Another interesting property of flux configurations is the amount of supersymmetry that they preserve. There exist in the literature several classes of supersymmetric fluxes (see e.g. \cite{gp}), but a unified understanding of them is lacking.
In this section we argue that our IIB flux configurations provide a
family of supersymmetric fluxes that interpolates between two classes
of familiar supersymmetric configurations \footnote{After completion of this work, we noticed \cite{fg} which discusses a
(potentially related) family of fluxes with interpolating supersymmetries. It
would be interesting to understand the relation between both approaches.}.

Indeed, considering the case of $\xi_6=\pm\xi_{10}\equiv \xi$, our flux background $G_3=F_3-\Phi H_3$ (with $\Phi$ the IIB complex coupling) satisfies the imaginary self- (or anti-self-)duality condition:
\beqa
G_3 = \pm i*_{6d} G_3
\eeqa
In fact, the fluxes can be seen to be $(2,1)$ [or $(1,2)$] and primitive.
Moreover the dilaton-axion fields are constant, and the metric takes
the form
\beqa
ds^2_{IIB} & = &
H_{5}^{-1/2}(H_5+\xi^2)^{1/2}ds^2_{0123}\, +\,
H_{5}^{1/2}(H_{5}+\xi^2)^{-1/2} \left \{  (dx^4)^2 +(dx^5)^2+\right. \nonumber \\ 
&&\left. + (H_{5}+\xi^2)ds^2_{789}+(H_{5}+\xi^2)^{-1}(dx^{6}+\omega)^2 \right \} 
\eeqa
It is a warped version of $\IM_4$ times $\IR^{2}_{(45)}$ times the Taub-NUT. This 
background is a particular case of the class considered in section 3 in \cite{gp} 
generalizing the fluxes in \cite{ks} (and which also appears in compact models, see 
e.g. \cite{gkp}).

On the other hand, for $\xi_6\neq\xi_{10}$, the flux $G_3$ does not have any 
particular self-duality property, the dilaton has a non-trivial profile, and the 
metric deviates from the warped ansatz. However, although the configuration is more 
involved, the M-theory picture implies that it is supersymmetric as well (preserving 
16 supercharges). On one hand the initial M5-brane configuration is clearly 
supersymmetric, and due to the smearing, the constant spinors do not depend on the 
duality coordinates 610, hence are also preserved in the dual IIB image \cite{tomas}. 
From a different point of view, the configuration satisfies the equations of motion, 
i.e. minimizes the energy, in a sector where the central charges satisfy a quadratic 
constraint, hence ensuring that the energy-minimizing state is 1/2 BPS, namely 
preserves 16 supersymmetries. It would be interesting to verify explicitly that the 
supersymmetric variations vanish in our background.

This more general class of supersymmetric fluxes, with varying dilaton, is highly 
reminiscent of the flux configurations appeared in \cite{ps} and studied in \cite{gp} 
(to first order in the fluxes). Indeed, the supersymmetries preserved correspond to 
the supersymmetry preserved by bound states of D3-branes and 5-branes. Moreover, as we 
discuss in section \ref{probes}, such fluxes induce mass terms and dielectric 
polarization on D3-brane probes when the latter are introduced, in close analogy to 
\cite{ps}. Again, it would be interesting to perform a direct comparison with the 
latter reference. Some differences exist, however: our configurations preserve 16 
supersymmetries, exist in Taub-NUT geometries, and are associated to $\NN=2$ mass 
terms on D3-brane probes (while fluxes in \cite{ps} preserve four supercharges, are 
introduced in (warped) flat space, and correspond to $\NN=1$ mass terms on D3-brane 
probes).

Leaving a more detailed discussion of this interesting property of our configurations 
for future work, we conclude by emphasizing that our duality provides a simple 
construction of a family of supersymmetric flux configurations interpolating between 
well-known classes of supersymmetric fluxes.

\subsubsection{Relation to magnetized D6-branes}
\label{magnetised}

Another interesting dual realization of the above systems is obtained as follows. 
Consider type IIB on a Taub-NUT geometry in 6789, momentarily without fluxes. Let us 
T-dualize along the direction 3, to obtain IIA theory on a Taub-NUT background. Now 
perform a `9-11 flip', namely lift the configuration to M-theory on a Taub-NUT 
background, and shrink the isometry direction of the latter; the resulting 
configuration is a type IIA D6-brane. Let us now carry out the same exercise in the 
presence of type IIB 3-form fluxes of the form (\ref{simpleflux}). T-dualizing to IIA 
we obtain a Taub-NUT geometry with fluxes
\beqa
H_3 = \xi_6\, \Omega \, dx^4 \quad ; \quad G_4 = \xi_{10}\, \Omega \, dx^3\, dx^5
\eeqa
Lifting to M-theory, we obtain a Taub-NUT geometry with 4-form flux
\beqa
G_4 = \xi_6\, \Omega \, dx^4 \, dx^{10} + \xi_{10}\, \Omega \, dx^3\, dx^5
\eeqa
This can be compared with fluxes in M-theory compactifications in \cite{beckers}. It 
is easy to recover different properties of the latter from our description in terms 
of M5-brane geometries.

Reducing along the isometrical direction of Taub-NUT, we obtain a IIA D6-brane. Now 
recall that D6-brane world-volume gauge fields arise from components of the M-theory 
3-form along the harmonic 2-forms, i.e. of the type $C_3=\Omega \, A_1$. We conclude 
that the configuration includes D6-brane world-volume magnetic fields
\beqa
F_2=\xi_6\, dx^4 \, dx^{10} + \xi_{10}\, dx^3\, dx^5
\eeqa
Namely there are constant magnetic fields $\xi_6$, $\xi_{10}$ in the 2-planes 410 and 
35, respectively. The imaginary self- (or anti-self-)duality conditions on the 
original IIB fluxes correspond to the (anti)self-duality of the world-volume gauge 
background. This dual picture provide interesting complementary viewpoints on our 
previous discussion and our proposals below.

\subsection{Multiple M5-branes and twisted fluxes on orbifolds}
\label{multiple}

\subsubsection{Relation to twisted fluxes in $\IC^2/\IZ_N$ orbifolds}
\label{twisted}

There is a natural generalization of the setup in section \ref{single}, namely 
considering several M5-branes. Consider the M-theory picture, and introduce $N$ 
M5-branes, labeled $a=1,\ldots, N$, spanning 0123 and the 2-planes
\beqa
x^6 & = & \tan \theta_{1,a} \, x^4 \nonumber \\
x^{10} & = & \tan \theta_{2,a}\, x^5
\eeqa
The dual IIB configuration is given by an 
$N$-center Taub-NUT space, with 3-form fluxes along the harmonic 2-forms
$\Omega_a$, namely
\beqa
H_3 & = & \sum_{a=1}^N \tan \theta_{1,a}\, \Omega_a \, dx^4 \nonumber \\
F_3 & = & \sum_{a=1}^N \tan \theta_{2,a}\, \Omega_a \, dx^5
\eeqa
Generalizations are straightforward and will not be discussed explicitly. It is 
interesting to realize that we in general have fluxes in the compact non-trivial 
2-cycles of the geometry. Namely, the projection of the flux along a particular 
harmonic 2-form $\Omega_{ab}$ is related to the relative angle between the 
corresponding dual M5-branes $a$, $b$. The M5-brane picture  allows a reliable 
description of the system even in the limit of coincident M5-branes, namely when 
the 2-cycles are collapsed to zero size and the geometry develops an orbifold 
singularity \footnote{Such fluxes have appeared e.g. in $\NN=2$ models in \cite{tt}.}. 
The present setup, therefore allows a reliable description of untwisted and twisted 
fluxes at orbifold singularities $\IC^2/\IZ_N$. The fact that the description and 
properties of the fluxes are reliable even in this large curvature regimes is 
clearly related to the large amount of supersymmetry preserved by the configuration.

\subsubsection{\bf Moduli stabilization and supersymmetry}
\label{moduli}

One of the most interesting features of configurations with fluxes is that 
minimization of the vacuum energy density leads to a constraint on the flux density, 
namely the flux combination $G_3=F_3-\Phi H_3$ (with $\Phi$ the IIB complex coupling) 
must be imaginary self-dual (or anti-self-dual)
\beqa
G_3=\pm i*_{6d}G_3
\label{isd} 
\eeqa
Since, due to flux quantization, flux densities are functions of moduli, the above 
condition fixes their vevs. In our particular context of fluxes in Taub-NUT geometries, 
the above condition is moreover equivalent to the conditions that the flux $G_3$ is 
$(2,1)$ [or (1,2)] and primitive (namely $G_3\wedge J=0$). That is, the flux preserves 
a particular supersymmetry. In this section we discuss how the above condition, moduli 
stabilization, and supersymmetry, are encoded in the M-theory configuration.

Consider compactifying the directions 45 in a two-torus, for simplicity rectangular, 
with $x^4$, $x^5$ of lengths $R_4$, $R_5$. Consider a configuration with two M5-branes, 
which we take without loss of generality at angles $\pm(\theta_1,\theta_2)$ in the 
two-planes 46 and 510, respectively. For simplicity, consider them to have wrapping 
numbers $k_6$, $k_{10}$ along 6, 10 (and wrapping once along 4, 5). We have
\beqa
\xi_6=\tan\theta_1=\frac{k_6R_6}{R_4} \quad ;\quad 
\xi_{10}=\tan \theta_2=\frac{k_{10}R_{10}}{R_5}
\eeqa
In the dual IIB picture we have a two-center Taub-NUT space with fluxes
\beqa
H_3  = 2 k_6 \, \Omega_{12}\, d\phi^4 \quad ; \quad
F_3  = 2 k_{10} \, \Omega_{12} \, d\phi^5
\eeqa
with $\Omega_{12}=(\Omega_1-\Omega_2)/2$ and $\phi^4,\phi^5$ are
defined as in section \ref{quantization}. Namely, it corresponds to a configuration 
with purely twisted fluxes.  

Introducing the complex coordinate $z=\phi^4+\tau \phi^5$ with $\tau=iR_5/R_4$, 
we have
\beqa
G_3\, =\,2\Omega_{12} \,\left [ \frac{dz}{\tau-{\ov\tau}} \, 
(\,k_{10}+\Phi\,{\ov\tau}\, k_6\,) \, + \, 
\frac{d{\ov z}}{\tau-{\ov\tau}} \, 
(\,-k_{10}-\Phi\,\tau\, k_6 \,) \right ]
\eeqa
The condition that $G_3$ is imaginary self-dual ($(2,1)$ and primitive) requires 
the last piece to drop, hence the dual of (\ref{isd}) is
\beqa
\frac{k_{10} R_{10}}{R_5}=\frac{k_6R_6}{R_4}
\eeqa

namely $\theta_1=\theta_2$. This is the familiar condition \footnote{The possibility 
$\theta_1=-\theta_2$ is dual to the (also energy minimizing) situation of $G_3$ being 
imaginary anti-self-dual fluxes (which in the Taub-NUT geometry is equivalent to being 
$(1,2)$ and primitive).} that the rotation relating the two M5-branes is in $SU(2)$. 
This is a condition of minimization of the energy of the two M5-brane system. When 
regarded as function of the moduli, the condition fixes the vevs of the latter, 
illustrating the dual version of moduli stabilization by fluxes. 

Finally, for configurations of branes intersecting at two non-trivial angles, the 
energy minimization conditions automatically imply the supersymmetry conditions. This is
dual to the statement that in Taub-NUT spaces, (anti)self-duality of the fluxes 
automatically implies supersymmetry of the configuration.

\subsubsection{Stabilization of non-susy flux configurations; enhancons}
\label{recombination}

Branes intersecting at angles {\em not} in $SU(2)$ relation suffer an instability 
against recombination to a single smooth curve, whose volume is smaller than the 
original intersecting configuration. This follows from analyzing the BPS formula 
for the system, or by using a dual intersecting D-brane configuration.
It is interesting to consider the interpretation of this process in the IIB dual 
realization of this system. In this section we discuss this interpretation in the 
case of none of the branes being along the directions 610 \footnote{This situation 
is considered in section \ref{dielectric}, and has a somewhat different dual 
interpretation, in terms of D3-brane polarization.}. A useful approach to this 
instability is to consider the supersymmetric case first, where the recombination 
process is parametrized by a modulus. Let $z=x^4+ix^5$, $w=x^6+ix^{10}$. The 
intersecting configuration is described by an M5-brane wrapped on a curve of the 
form $(z-w)(z+w)=0$, while in the recombined configuration the brane wraps the cycle 
$(z-w)(z+w)=\epsilon$.

Consider as in section \ref{moduli} two M5-branes at angles $\pm (\theta_1, \theta_2)$ 
in the 46 and 510 directions, respectively, with wrapping numbers $(1,k_6)$ and 
$(1,k_{10})$, and which coincide in 789. The dual configuration contains a two-center 
Taub-NUT geometry, with coincident center, and with 3-form twisted fluxes 
\beqa
H_3 = 2\,\xi_6\, \Omega_{12}\, dx^4 \quad ; \quad 
F_3 = 2\,\xi_{10}\, \Omega_{12}\, dx^5
\eeqa
For $\xi_6\neq\xi_{10}$, the flux configuration is non-supersymmetric, and the 
M-theory picture suggests that it suffers an instability. 

The process involves somewhat exotic degrees of freedom. In the supersymmetric case, 
the degrees of freedom responsible for the restabilization of the configuration are 
BPS states, and so can be translated directly from dual configurations of intersecting 
M5- or D-branes. It is easy to realize that they correspond to tensionless degrees of 
freedom from D3-branes wrapped on the collapsed 2-cycle $\Sigma_{12}$. D3-branes 
wrapped on the $\Sigma_{12}$ lead to tensionless objects because there are points in 
45 where the integral of the 2-form fields over $\Sigma_{12}$ vanish. The number of 
such points, and hence of the massless degrees of freedom involved in the condensation, 
is given by $4k_6k_{10}$, in agreement with the number of intersection points in the 
dual.

The nature of the configuration after condensation of these degrees of freedom is 
unclear, since it involves a Higgs effect for the 2-form fields. It is tantalizing to 
propose that its description in the IIB configuration is related to the enhancon 
geometries \cite{enhancon}, since it involves a shell region (dual to the region of 
recombination of the dual M5-branes) associated to massless D-brane degrees of 
freedom. Indeed, in M-theory the final configuration is an M5-brane wrapped on a 
single smooth curve. The IIB version of this kind of configuration in theories with 
8 supercharges has been discussed in \cite{polchitwo}. It would be very interesting 
to develop these relations further.

Although the decay involves a configuration not very familiar in the context of fluxes, 
it is perfectly natural in other dual realizations of the system. Consider for instance 
the dual configuration in section \ref{magnetised}, of D6-branes with world-volume 
magnetic fluxes. In the non-supersymmetric case the abelian gauge background decays to 
an $U(2)$ instanton involving the non-abelian degrees of freedom, arising from open 
strings stretching between the two branes. The non-abelian character of the final 
configuration is thus responsible for the difficulty in finding a supergravity 
description on the IIB side. 

Note that the above argument provides and answer to the question raised in section 
\ref{induced} on the IIB dual to M5-branes wrapped on non-factorized holomorphic 
cycles. Interestingly, the latter provide a geometric picture of a somewhat mysterious 
process in IIB theory.

\subsubsection{Stabilization/supersymmetrization of non-susy geometries}
\label{tachyons}

There is an interesting dual realization of this kind of configuration, in which one
of the angle parameters maps to a flux background, while the second angle leads
to a non-Calabi-Yau geometry in the dual. The full configuration is however stable
and supersymmetric, due to a compensating effect of the flux and the geometry. 

Consider two M5 branes spanning 0123, at angles $\pm (\theta_1, \theta_2)$ in 
the 2-planes 46 and 510, and located at points in 789. Consider the 
coordinate 9 to be compact, and consider reducing to a type IIA configuration 
by shrinking it. We obtain two IIA NS5-branes spanning 0123, at angles 
$\theta_1$ in 46 and $\theta_2$ in 510. Let us now perform a T-duality 
along the direction 10. In the case of $\theta_2=0$, the T-dual configuration 
is a purely geometric background that corresponds to a non-supersymmetric 
and unstable background, described by a 5d geometry $\IX_5$, considered 
in \cite{coninst}. Intuitively, it corresponds to two Taub-NUTs 
intersecting in a non-supersymmetric fashion. The configuration contains 
an instability localized at the singular point at the intersection. The 
relaxation of the instability was shown to correspond to a dynamical resolution 
of the singularity, leading to a smooth geometry with a non-trivial 2-sphere, which 
in the non-compact case runs away to infinite size.

For non-zero $\theta_2$, the new non-trivial angle does not modify the topology 
dual geometry. The IIB configuration is given by a squashed version of the same
topological manifold $\IX_5$, which in addition now contains a non-zero NSNS 
3-form flux. 

For the case $\theta_1=\pm \theta_2$, namely for a 3-form flux tuned to the 5d 
geometry, the configuration is supersymmetric. In this sense, the introduction of the 
3-form flux stabilizes the geometry $\IX_5$. In fact, the unstable direction, which 
corresponds to increasing the size of the non-trivial 2-sphere, is a modulus direction 
(dual to the recombination of the two intersecting branes in the susy case). It is
interesting to notice that in the large volume limit of this modulus, supergravity is
reliable and the stabilization mechanism is amenable to explicit analysis.

Clearly $SL(2,\IR)$ transformations of the above configuration (equivalently,
shrinking the directions 910 in different fashions) show that the
stabilization can be achieved for a suitably tuned amount of RR flux or a
combination of NSNS and RR fluxes.

Although the above non-supersymmetric geometries are not of orbifold kind, it is 
tempting to speculate that flux configurations in non-supersymmetric orbifold may also 
lead to the stabilization of such geometries (although perhaps not in a supersymmetric 
fashion). This presumably arises from the competition of the tachyonic nature of 
twisted sector scalars and the contribution to their potential generated by a 
background of twisted fluxes. It would be interesting to analyze this idea in more 
detail.

\subsubsection{\bf Dualities to conifolds}
\label{dualconi}

Finally, we would like to point out that the above configurations admit a dual version
where both angles become properties of the geometry; indeed, the configurations are 
dual to a purely geometric background containing a conifold singularity. Consider two 
M5-branes intersecting at $SU(2)$ angles \footnote{The duality for non-supersymmetric 
angles can be performed similarly, and leads to non-supersymmetric geometries of 
conifold type, considered in \cite{coninst}.}, and consider compactifying, and 
subsequently shrinking, the directions 78. For instance, shrinking along 8, we 
obtain a configuration of intersecting IIA NS5-branes; and T-dualizing along 7, 
we obtain \cite{uraconi} IIB theory on a singular geometric background with a conifold 
singularity. 

Notice that the two dualities are related in the M-theory picture by an exchange of 
the directions 67 and 810, namely a simultaneous $SL(2,\IZ)$ transformation in the 
corresponding two two-tori. In the IIB picture, the configuration of Taub-NUT with 
fluxes and the conifold geometry are thus related by a U-duality transformation. It 
is clear that one can generate even more duals by choosing to dualize along other 
intermediate directions in the two-tori 67 and 810. All these would fill out a 
multiplet of highly non-trivial configurations under U-duality.

\subsection{D-brane probes}
\label{probes}

In this section we consider several properties of the physics of D3-brane probes
in the above Taub-NUT plus flux configurations, gaining insight from the dual
M-theory realization. In the latter, the D3-branes correspond to M5-branes
along 0123610. 

\subsubsection{Mass deformations}
\label{masses}

The M-theory configurations we are considering have an interesting application.
Configurations of M5-branes (reducing to NS5- and D4-branes in IIA) have been 
considered in \cite{witten} to study the dynamics of the $\NN=2$ 4d supersymmetric 
gauge theories on the non-compact part of their world-volume. In particular, the 
situation with one M5-brane along 012345 and $N$ M5-branes along 0123610 leads to 
an $\NN=2$ $U(N)$ gauge theory with an adjoint hypermultiplet (that is, $\NN=4$ 
super-Yang-Mills).

\begin{figure}
\begin{center}
\centering
\epsfysize=3.5cm
\leavevmode
\epsfbox{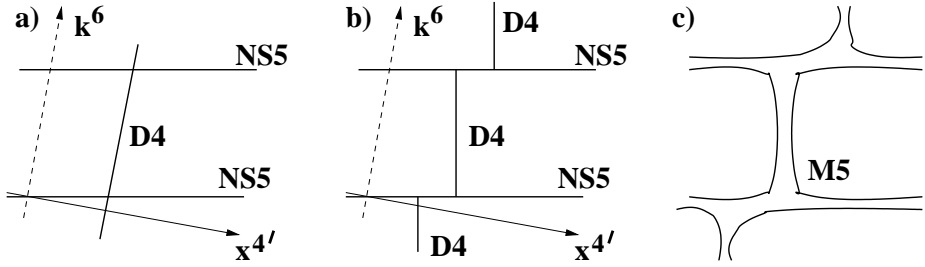}
\end{center}
\caption[]{\small The type IIA picture dual to IIB D3-branes in the presence of flux 
corresponds to NS5/D4 brane configurations in a twisted $\IS^1\times \IR$. a) The 
duals to D3-branes (wrapped D4-branes) are attracted (in 789) towards the duals of the 
Taub-NUT center (NS5-branes). b) Once there, they break on the NS5-branes and relax to 
a supersymmetric configuration; in the T-dual this corresponds to a dielectric 
polarization. c) The M-theory picture of b) is a recombination of initially 
intersecting M5-branes.}
\label{massdef}
\end{figure}

An interesting deformation of this theory corresponds to the introduction of a complex 
mass parameter for the hypermultiplet. In the brane realization, it was argued in 
\cite{witten} to correspond to embedding the M5-brane configuration in a non-trivial 
geometry in 45610, defined by a twisted identification (for diagonal metric in 610)
\beqa
x^6\to x^6+2\pi R_6 \quad ; \quad v=x^4+ix^5 \to v+m
\label{twist}
\eeqa
In the IIA picture, when D4-branes along 01236 are introduced, the shift implies that 
it is energetically favourable for them to break at the NS5-brane location, see figure 
\ref{massdef} a, b. The separation $m$ between the D4-brane is translated into a 
hypermultiplet mass. In the M-theory picture, the breaking corresponds to a 
recombination of the different kinds of M5-branes, figure \ref{massdef}c, which due to 
the twisting no longer intersect in $SU(2)$ angles. The holomorphic curve wrapped by 
the recombined M5-brane after recombination is the Seiberg-Witten curve of the 
mass-deformed theory.

It is easy to realize that the configuration of the M5-brane along 012345 in the 
twisted geometry (\ref{twist}) is one particular case of our M-theory 
configurations. Indeed, consider one M5-brane along 012345, and introduce the variables
\beqa
x^4=x'^4-\frac{Re(m)}{2\pi R_6}x^6 \quad ; \quad 
x^5=x'^5-\frac{Im(m)}{2\pi R_6}x^6
\eeqa
We see that at fixed values of the coordinates $x^4$, $x^5$, there is an induced 
shift of $x'^4$, $x'^5$ when $x^6$ is shifted, precisely given by (\ref{twist}).
Hence, we reach the conclusion that our configurations provide a realization of the 
mass deformation of $\NN=4$ super-Yang-Mills, once suitable probes are introduced. It 
is useful to see how this arises in the dual IIB picture. We have a configuration of 
$N$ D3-branes in a Taub-NUT geometry with $H_3$ flux roughly of the form $H_3= \Omega 
\wedge (m_4 dx^4 + m_5 dx^5)$. The configuration does not preserve supersymmetry, and 
the D3-branes suffer an attractive force that pins them to the Taub-NUT center. The 
pinning potential for the adjoint hypermultiplet, has been discussed in a dual 
realization in \cite{pinned}. The mass term is as expected given by $m$.

A last tricky point is restoration of supersymmetry, which is provided by the T-dual 
process of the D4-brane breaking on the NS5-brane. This effect will be considered in 
more detail in next section, where it is shown to correspond to a dielectric 
polarization of the D3-branes, as in \cite{myers}. Indeed this is suggested by the 
fact that the two endpoints of the D4-brane are oppositely charged under the NS5-brane 
world-volume fields, and they are separated (in a dipole-like fashion) due to the 
twisted geometry. 

\subsubsection{Myers dielectric effect}
\label{dielectric}

In this section we consider the main dynamical process taking place in certain non-supersymmetric configurations. Consider one M5-brane at angles $\theta_1$, $\theta_2$ in the 2-planes 46 and 510, generically with $\theta_1\neq \pm \theta_2$. Let us now introduce M5-branes spanning 0123610; these do not preserve the
same supersymmetry as the original ones. The configuration suffers from an instability against recombination of the M5-branes. Once they recombine, we end up with a single M5-brane wrapping a supersymmetric cycle, preserving a supersymmetry different from any of the original ones.

In the dual IIB configuration, we have a Taub-NUT geometry with 3-form flux $G_3$, generically not satisfying any (anti)self-duality condition. In this background we introduce a set of D3-brane probes. In general, they do not preserve the same supersymmetry as the background fluxes. The M-theory picture suggests there is an instability in this situation, and a mechanism that allows to restabilize the configuration and restore the supersymmetry.
Indeed, there is such a process, as is suggested by the dielectric effect \cite{myers}. The world-volume action for a D3-brane in a non imaginary self-dual flux background develops trilinear couplings for the world-volume scalars $\phi_m$
\beqa
(G_3-i*_{6d}G_3)_{lmn} \,\phi_l\, \phi_m\,\phi_n
\eeqa
which trigger polarization of the D3-brane into a fuzzy sphere \cite{myers}. It is non-trivial to compute the final endpoint of the relaxation process, including the backreaction of the dielectric D3-brane on the background. Happily the M-theory picture provides the answer and ensures that the final state is supersymmetric. 

Notice the close analogy of the above discussion with the picture in \cite{ps}. Namely, a set of D3-branes immersed in a supersymmetric but non-imaginary-selfdual flux is polarized to a final configuration (including the backreaction) which preserves as a whole some supersymmetry. Moreover, in \cite{ps} the fluxes correspond to a deformation of the D3-brane world-volume theory by mass terms for the adjoint matter, precisely as in our case, as
discussed in previous section. The main difference between both configurations is that in \cite{ps} the configurations preserve four supercharges, and correspond to mass terms for all $\NN=1$ chiral multiplets, while in our case the configuration preserves eight supercharges, and the fluxes correspond to an adjoint hypermultiplet mass \footnote{A further difference is that in
our configuration the background geometry is Taub-NUT, rather than flat space. Hence a full comparison would require taking a decompactification limit, where Taub-NUT reduces to flat space.}.
It would be very interesting to compare our backgrounds with the AdS/CFT duals proposed for $\NN=2$ mass deformed theories in e.g. \cite{warner}.

\section{Conifolds and $N=1$ fluxes}
\label{foursusy}

It is natural to wonder if the above duality between fluxes and brane configurations 
admits a generalization to background geometries different from Taub-NUT, for instance 
the interesting case of Calabi-Yau threefolds. Clearly, our duality has been useful
because the background geometry could be dualized into a brane, which subsequently 
accounts for the background fluxes by a deformation of its geometry. In order to be
able to repeat the exercise for some Calabi-Yau threefold, we need the corresponding 
geometry to be related to configurations of branes as well. Namely, we should look for
an underlying duality between M5-branes and threefold geometries, analogous to
that in section \ref{dualityone}.

A natural candidate is the duality between a configuration of two M5-branes along 
012345 and 012389, respectively, and type IIB theory on a threefold geometry $\IX_6$ 
containing a conifold  singularity \cite{uraconi}. An intuitive derivation is as 
follows. Let us start in the M-theory picture with 6, 10 compact, and M5-branes along 
012345 and 012389, and let us find the dual IIB picture when we shrink 6, 10. For 
instance, shrinking 10 first we get a IIA configuration of two NS5-branes along 012345
and 012389. Denoting $z$, $w$ the complex coordinates in the two 2-planes 45 and 89,
we have an NS5-brane wrapped on the curve $zw=0$. Let us now perform a T-duality along
the direction 6. The T-dual configuration must be a purely geometric background given
by `two intersecting Taub-NUTs', namely an $\IS^1$ fibration (with fiber parametrized
by the dual $x^6$), degenerating over the curved wrapped by the original NS5, i.e. over 
the locus $zw=0$. The fibration can be described by the equation
\beqa
xy=zw
\eeqa
where the two complex variables $x$, $y$, subject to the above constraint, encode the 
coordinates 67. Namely, for generic $z$, $w$, the variables $x$, $y$ parametrize a 
$\IC^*$, that is a cylinder; at the locus $zw=0$ the circle degenerates to zero size, 
as desired.
The above equation defines a conifold singularity, in the complex sense. The precise
metric arising from T-duality would differ from the standard conifold metric in that
our geometry has constant asymptotic radius for the fiber, while the standard conifold
is a conical singularity where all directions grow to infinity with the radius. However,
both metrics agree near the origin; in fact the duality has been worked out using 
the (suitably smeared) M5-brane supergravity solution and the T-duality rules, 
showing that the near throat region of the M5-branes dualizes into a standard conifold 
singularity (see second reference in \cite{uraconi}). We will not enter into the 
detailed analysis.

There are two branches in the moduli space of the M5-brane configuration: 
the two M5-branes can be 
separated in the directions 67, or if they coincide in 67 they can be recombined
(by an amount fixed by a complex modulus $\epsilon$) into a single smooth M5 brane, 
wrapped on the curve $zw=\epsilon$. These two branches can be intuitively T-dualized
into the small resolution of the conifold (with a homologically non-trivial $\IS^2$,
with size and $B$-field fixed by the interbrane distance in 76), and into the 
deformation of the conifold (with a homologically  non-trivial $\IS^3$, with size
fixed by $\epsilon$), described by a complex manifold $xy-zw=\epsilon$.

\medskip

There is a notable example of configuration of fluxes in IIB theory on the (deformed) 
conifold which has been studied in the literature, namely the Klebanov-Strassler 
solution \cite{ks}, proposed as a supergravity dual of the theory of fractional 
D3-branes on the conifold. In the remainder of this section we would like to argue 
that our duality relates this background with a configuration of M5-branes wrapped 
on a holomorphic curve in flat space. The latter is the M-theory lift of a IIA 
configuration with NS5-branes along 012345 and 012389, with D4-brane  suspended among 
them. In fact, such a duality was already noticed in \cite{ks} and studied e.g. in 
\cite{tatar}. In this section we review those results, pointing out that they can be 
understood in a broader context as a generalization of our previous duality for 
Taub-NUT spaces.

The curve represents two M5-branes along 012345 and 012389 recombined together via a 
piece of M5-brane along 0123610. The curve has been determined in \cite{wittenone} 
(for the case of non-compact 6, which is enough for qualitative purposes), and is 
given by
\beqa
z=\epsilon w^{-1} \quad ; \quad t=w^N
\eeqa
where $z$ and $w$ parametrize 45 and 89, $t$ parametrizes 6, 10, and $\epsilon$ is
related to the amount of recombination of the branes. Its holomorphy
is a reflection of the supersymmetry of the system. 

\medskip

In the present setup, we see that the proposed duality satisfies several points in 
analogy with the previous Taub-NUT examples. For instance, in the M5-brane
configuration, the coordinates $z$, $w$ satisfy $zw=\epsilon$. Hence the dual IIB
background geometry should correspond with a deformed conifold, exactly as in \cite{ks}
\footnote{Moreover, in both pictures the appearance of the scale $\epsilon$ is
associated to strong dynamics in the low-energy gauge field theory.}. 

The fact that the M5-brane does not sit at fixed positions in $t$ (i.e. in 610) implies
the dual IIB configuration contains non-trivial $F_3$ and $H_3$ fluxes. In principle
these fluxes would be turned on along harmonic forms associated to the two intersecting
Taub-NUT spaces implicit in the conifold geometry. However, by taking linear 
combinations, the fluxes can be seen be turned along the non-trivial $\IS^3$ and its
dual. Using $SL(2,\IZ)$ transformations, the flux over $\IS^3$ can be chosen to be 
$H_3$, while (as we argue shortly) the flux on its dual turns out to be $F_3$.

Holomorphy of the curve implies that $t$ (or 610) depends holomorphically on the 
coordinates of the base $z$, $w$. As we know this implies that dual flux configuration 
is supersymmetric, namely $G_3$ is $(2,1)$ and primitive, precisely as in \cite{ks}. 
Since this implies imaginary self-duality of the flux, it implies that fluxes along 
the $\IS^3$ and its dual can be chosen to be $H_3$, $F_3$.

Finally, the amount of M5-brane spanning in 610 is mapped in the dual to the integral 
of $F_3\wedge H_3$ in the dual IIB configuration. Namely, similarly to our Taub-NUT
examples, the dual background reproduces the 4-form charge purely in terms of fluxes, 
as in \cite{ks}.

\medskip

The present duality admits clear generalizations to other threefolds having a simple
dual in terms of M5-brane configurations. Several examples of this kind have been
considered in the first reference in \cite{uraconi}, for instance threefolds given by 
equations $xy=z^Nw^M$ are dual to configurations of $N$ and $M$ M5-branes along 012345 
and 012389. Unfortunately the metrics for these manifolds are not known, even in the 
near singularity region. We hope that a more quantitative understanding of our duality
in the threefold case can lead to a more detailed understanding of fluxes in these
geometries, hopefully providing new examples useful e.g. for the construction of 
supergravity duals of gauge theories.

\section{Conclusions}
\label{conclusions}
In this paper we have exploited duality to relate backgrounds of 3-form field strength 
fluxes in IIB string theory on Taub-NUT space with configurations of M-theory 
M5-branes. Properties of the latter are familiar, yet can be related to interesting 
old and new properties of the flux configurations. Among the old properties, we have 
discussed flux quantization, supersymmetry conditions and moduli stabilization for 
fluxes, from the dual M5-brane geometry. We have also uncovered new aspects, like the 
possibility of stabilizing non-supersymmetric unstable compactifications by combining 
them with fluxes. Among the new properties, we have provided dual pictures for flux 
configurations in complicated (large curvature) regimes, like twisted fluxes on 
$\IC^2/\IZ_N$ orbifolds, and for complicated dynamical processes in the flux 
configurations, like restabilization of non-supersymmetric fluxes, and Myers 
dielectric effect. 

These results indicate that duality properties of configurations with fluxes can be 
extremely useful in improving their understanding. A related direction, initiated in 
the study of mirror symmetry with fluxes \cite{mirror}, would be to understand duality 
properties of fluxes in compact manifolds. Indeed, we hope much active research along 
these lines.

\medskip

\centerline{\bf Acknowledgements}

We thank the Theoretical Physics group at the University of Hamburg, and particularly 
Jan Louis, for their hospitality during completion of this work. A.M.U. thanks 
M.~Gonz\'alez for kind encouragement and support. J.G.C wants to thank M. P\'erez for 
her patience and affection. This work has been partially supported by CICYT (Spain). 
The research of J.G.C. is supported by the Ministerio de Educaci\'on, Cultura y 
Deporte through a FPU grant. 

\newpage

\appendix

\section{Buscher's rules for T-duality}
\label{trules}

The relations between background fields between IIA and IIB configurations related by 
T-duality along the direction $y$ are given by (see e.g. \cite{om}). 
\beqa
\tilde{G}_{ yy} &=& {1\over G_{yy}}
\qquad\qquad\qquad\qquad
\qquad\qquad\qquad\qquad
e^{2 \tilde{\phi}} =\, { e^{2 \phi} \over G_{yy}}
\nonumber\\
\tilde{G}_{ \mu \nu} &=& G_{\mu \nu}
- { G_{\mu y} G_{\nu y}
- B_{\mu y} B_{\nu y}
\over G_{ yy}}
\qquad\qquad\qquad\quad
\tilde{G}_{\mu y} ={ B_{\mu y}
\over G_{yy}}
\label{NSrule}\\
\tilde{B}_{ \mu \nu}&=&B_{ \mu \nu}
-{B_{\mu y} G_{\nu y}-G_{\mu y}
B_{\nu y}\over G_{yy}}
\qquad\qquad\qquad\quad
\tilde{B}_{\mu y} ={ G_{\mu y}
\over G_{ yy}}
\nonumber
\eeqa             

\beqa
\tilde{C}^{(n)}_{\mu\cdots\nu\alpha y}&=&
C^{(n-1)}_{\mu\cdots\nu\alpha}-(n-1)
{C^{(n-1)}_{[\mu\cdots\nu| y}G_{|\alpha]y}\over G_{yy}}
\label{RRrule}\\
\tilde{C}^{(n)}_{\mu\cdots\nu\alpha\beta}&=&
C^{(n+1)}_{\mu\cdots\nu\alpha\beta y}
+nC^{(n-1)}_{[\mu\cdots\nu\alpha}B_{\beta]y}
+n(n-1){C^{(n-1)}_{[\mu\cdots\nu|y}B_{|\alpha|y}G_{|\beta]y}\over
G_{yy}}
\nonumber
\eeqa

\end{document}